\documentclass{article}
\usepackage{spconf,amsmath,amsfonts,graphicx,booktabs}
\usepackage[dvipsnames]{xcolor}
\usepackage{diagbox}
\usepackage{multirow}
\usepackage[linktocpage=true,hyperfootnotes=false]{hyperref}

\usepackage{caption}
\usepackage{subcaption}
\usepackage{cite}

\usepackage{listings}
\usepackage{xcolor}

\usepackage{amssymb}
\usepackage{pifont}
\newcommand{\cmark}{\ding{51}}%

\definecolor{codegreen}{rgb}{0,0.6,0}
\definecolor{codegray}{rgb}{0.5,0.5,0.5}
\definecolor{codepurple}{rgb}{0.58,0,0.82}
\definecolor{backcolour}{rgb}{0.95,0.95,0.92}

\lstdefinestyle{mystyle}{
	backgroundcolor=\color{backcolour},   
	commentstyle=\color{codegreen},
	keywordstyle=\color{magenta},
	numberstyle=\tiny\color{codegray},
	stringstyle=\color{codepurple},
	basicstyle=\ttfamily\footnotesize,
	breakatwhitespace=false,         
	breaklines=true,                 
	captionpos=b,                    
	keepspaces=true,                 
	numbers=left,                    
	numbersep=5pt,                  
	showspaces=false,                
	showstringspaces=false,
	showtabs=false,                  
	tabsize=2
}

\usepackage{enumitem}
\setlist[itemize]{leftmargin=*}
\setlist[enumerate]{leftmargin=*}

\usepackage{color}

\usepackage{scrextend}

\title{\vspace{-7mm}Adversarial permutation invariant training\\for universal sound separation}
\name{Emilian Postolache\sthanks{Equal contribution.}$^{1,2}$, Jordi Pons$^{\ast1}$, Santiago Pascual$^1$, Joan Serrà$^1$\vspace{-2mm}}

\address{$^1$Dolby Laboratories \hspace{10mm} $^2$Sapienza University of Rome\vspace{-3mm}}

\begin{document}
	\ninept
	\maketitle

	\begin{abstract} 
			\vspace{1mm}
		Universal sound separation consists of separating mixes with arbitrary sounds of different types, and permutation invariant training~(PIT) is used to train source agnostic models that do so. In this work, we complement PIT with adversarial losses but find it challenging with the standard formulation used in speech source separation. We overcome this challenge with a novel $I$-replacement context-based adversarial loss, and by training with multiple discriminators.~{Our experiments show that by simply improving the loss}  (keeping the same model and dataset) we obtain a non-negligible improvement of 1.4\,dB $\text{SI-SNR}_\text{I}$ in the reverberant FUSS dataset. We also find adversarial PIT to be effective at reducing spectral holes, ubiquitous in mask-based separation models, which highlights {the potential relevance of adversarial losses for source separation.}
	\end{abstract}

	\section{Introduction}

	Audio source separation consists of separating the sources present in an audio mix, as in music source separation (separating vocals, bass, and drums from a music mix~\cite{stoller2018adversarial,lluis2019end, pons2020upsampling}) or speech source separation (separating various speakers talking simultaneously~\cite{kadiouglu2020empirical, liu2021permutation,narayanaswamy2020unsupervised}).
	Recently, universal sound separation was proposed~\cite{kavalerov2019universal}. It consists of building {source agnostic} models that are not constrained to a specific domain (like music or speech) and can separate any source given an arbitrary mix. 
	Permutation invariant training~(PIT)~\cite{yu2017permutation} is used for training universal source separation models based on deep learning~\cite{kavalerov2019universal,wisdom2021s,wisdom2020unsupervised}. 
	
	We consider mixes $m$ of length $L$ with $K'$ arbitrary sources $\mathbf{{s}}$ as follows: $m = \sum_{k=1}^{K'} s_k$, out of which the separator model $f_\mathbf{\theta}$ predicts $K$ sources $\mathbf{\hat{s}} = f_\mathbf{\theta}(m)$. 
	PIT optimizes the learnable parameters $\mathbf{\theta}$ of $f_\mathbf{\theta}$ by minimizing the following permutation invariant loss:
	
	\vspace{-1.5mm}
	
	\begin{equation}
		\label{eq:pit}
		\mathcal{L}_{\text{PIT}} = \min_{\mathbf{P}} \sum_{k=1}^{K}  \mathcal{L}\left(s_k,  \left[\mathbf{P}{\hat{\mathbf{s}}}\right]_k\right),
	\end{equation}

	\noindent where we consider all permutation matrices $\mathbf{P}$, $\mathbf{P}^*$ is the optimal permutation matrix minimizing Eq.~\ref{eq:pit}, and $\mathcal{L}$ can be any regression loss. 
	Since $f_\mathbf{\theta}$ outputs $K$ sources, in case a mix contains $K'<K$ sources, we set the target ${s}_k=0$ for $k>K'$. 
	Note that a permutation invariant loss is required to build {source agnostic} models, because the outputs of $f_\mathbf{\theta}$ can be any source and in any order. 
	As such, the model must not focus on predicting one source type per output, and any possible permutation of output sources must be equally correct~\cite{kavalerov2019universal,yu2017permutation}. 
	A common loss $\mathcal{L}$ for universal sound separation is the $\tau$-thresholded logarithmic mean squared error~\cite{kavalerov2019universal,wisdom2021s}, which is unbounded when ${s}_k=0$. In that case, since $m\ne0$, one can use a different $\mathcal{L}$ based on thresholding with respect to the mixture~\cite{wisdom2021s}:
	\begin{align}
		\label{eq:loss}	
		\mathcal{L}(s_k,\hat{s}_k) = \begin{cases}
			10 \log_{10}\left(\Vert \hat{{s}_k} \Vert^2 + \tau \Vert m \Vert^2\right) & \text{if } s_k = 0 \\ 10 \log_{10}\left(\Vert {s}_k - \hat{{s}}_k \Vert^2 + \tau \Vert {s}_k \Vert^2\right) & \text{otherwise.}
		\end{cases}
	\end{align}

	\begin{table*}[h!]
		\resizebox{\textwidth}{!}{%
			\begin{tabular}{l c c c c c}
				
				\toprule
				
				\textbf{Previous works on (two speaker)}& \textit{Discriminator type and input} & \multirow{2}{*}{\textit{Multiple $D$?}} & \textit{Input to $D$} &  \textit{ With $\mathcal{L}_{\text{PIT}}$?} &\textit{Adversarial}\\ 
				{\textbf{speech source separation}}& {\textit{(Type $\rightarrow$ Input: real / fake)}}  &  & \textit{(domain)} & \textit{(+$\mathcal{L}_{\text{PIT}}$ domain)} & \textit{loss} \\ \midrule
				
				CBLDNN \cite{li2018cbldnn} &{$D_{\text{ctx}, I=0}^\text{STFT}$ \hspace{1mm}$\rightarrow$\hspace{1mm} $[m, {s_1}, {s_2}]$ / $[m, \hat{s}_1, \hat{s}_2]$ ($m$-conditioned) }& No & $\vert$STFT$\vert$ & Yes, $\vert$STFT$\vert$ & LSGAN \\			
				SSGAN-PIT \cite{chen2018permutation}: variant (i) & {$D_{\text{ctx}, I=0}^\text{STFT}$ \hspace{1mm}$\rightarrow$\hspace{1mm} $[m, {s_1}, {s_2}]$ / $[m, \hat{s}_1, \hat{s}_2]$ ($m$-conditioned)} & No & $\vert$STFT$\vert$ & Yes, $\vert$STFT$\vert$ & LSGAN  \\ 
				SSGAN-PIT \cite{chen2018permutation}: variant (ii)  & $D_{\text{ctx}, I=0}^\text{STFT}$ \hspace{1mm}$\rightarrow$\hspace{1mm} $[{s_1}, {s_2}]$ / $[\hat{s}_1, \hat{s}_2]$ & No & $\vert$STFT$\vert$ &  Yes, $\vert$STFT$\vert$ & LSGAN \\ 
				SSGAN-PIT \cite{chen2018permutation}: variant (iii)  & $D_{\text{inst}}^\text{STFT}$ \hspace{1mm}$\rightarrow$\hspace{1mm} $[{s}_1]$ / $[\hat{s}_1]$ \quad and \quad $[{s}_2]$ / $[\hat{s}_2]$& No & $\vert$STFT$\vert$ & Yes, $\vert$STFT$\vert$ & LSGAN  \\ 	
				Furcax \cite{shi2019furcax} &$D_{\text{ctx}, I=0}^\text{wave}$ \hspace{1mm}$\rightarrow$\hspace{1mm} $[{s_1}, {s_2}]$ / $[\hat{s}_1, \hat{s}_2]$ & No & Waveform & Yes, waveform & LSGAN \\  				
				Conv-TasSAN \cite{deng2020conv} & Metric predictor \hspace{1mm}$\rightarrow$\hspace{1mm} $[{s_1}, {s_2}, {s_1}, {s_2}]$ / $[\hat{s}_1, \hat{s}_2, {s_1}, {s_2}]$  & No & Waveform & Yes, waveform & MetricGAN \\ \midrule
				
			 \multirow{3}{*}{\shortstack[l]{\textbf{Our source agnostic method} \\ \textbf{for universal sound separation}}} & { Any above except ``metric predictor''; with more than} & \multirow{3}{*}{\shortstack[c]{{Yes, for} \\ {better quality}}} & \multirow{3}{*}{\shortstack[c]{{Any above,} \\ {plus masks}}}  &  & \\ 
				& {two input sources; using ${D}_{\text{ctx,}I> 0}$ with $I$-replacement;} & &  & Optional &  Hinge loss  \\
				& {and with source agnostic discriminators } &  &  & &   \\ 
				\bottomrule	
			\end{tabular}
		}
		\vspace{-1mm}
		\caption{Summary of differences between our work (bottom) and previous adversarial PIT works for speech source separation (top).}
		\vspace{-2mm}
		\label{table:sota}
	\end{table*}

	\noindent In this work, we complement PIT with adversarial losses for universal sound separation. 
	A number of speech source separation works also complemented PIT with adversarial losses~\cite{li2018cbldnn,chen2018permutation,shi2019furcax,deng2020conv}. Yet, we find that the adversarial PIT formulation used in speech separation does not perform well for universal source separation (sections~\ref{setup} and~\ref{discussion}). To improve upon that, in section~\ref{adversarialPIT} we extend speech separation works with: a novel $I$-replacement context-based adversarial loss, by combining multiple discriminators, and generalize adversarial PIT such that it works for universal sound separation (with source agnostic discriminators dealing with more than two sources). Table~\ref{table:sota} outlines how our approach compares with speech separation works.

	\section{Adversarial PIT}
	\label{adversarialPIT}

	Adversarial training, in the context of source separation, consists of simultaneously training two models: $f_\theta$ producing plausible separations $\mathbf{\hat{s}}$, and one (or multiple) discriminator(s) $D$ assessing if separations $\mathbf{\hat{s}}$ are produced by $f_\theta$ (fake) or are ground-truth separations $\mathbf{{s}}$ (real). 
	Under this setup, the goal of $f_\theta$ is to estimate (fake) separations that are as close as possible to the (real) ones from the dataset, such that $D$ misclassifies \mbox{$\mathbf{\hat{s}}$ as $\mathbf{{s}}$}~\cite{goodfellow2020generative, pascual2017segan}. 
	We propose combining variations of an instance-based discriminator~$D_{\text{inst}}$ with a novel $I$-replacement context-based discriminator~$D_{\text{ctx}, I}$. 
	Each $D$ has a different role and is applicable to various domains: waveforms, magnitude STFTs, or masks.
	Without loss of generality, we present $D_{\text{inst}}$ and $D_{\text{ctx}, I}$ in the waveform domain and then show how to combine multiple discriminators operating at various domains to train $f_\theta$.

\begin{figure}[t]
		\centering
		\vspace{-1mm}
		\hspace{8mm}\includegraphics[width=0.6\linewidth]{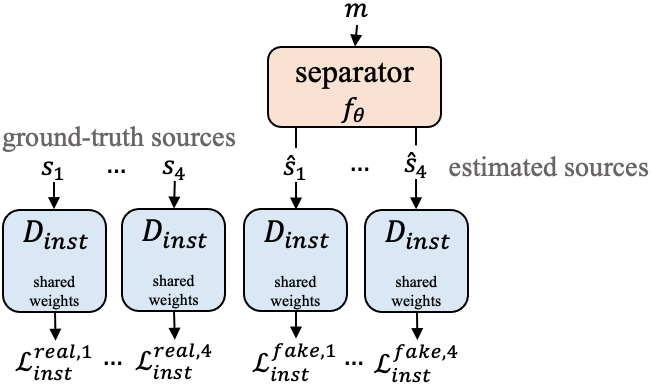}
		\caption{Instance-based adversarial loss ($K = 4$). }
		\label{fig:instance}
		\vspace{-3mm}
	\end{figure}

	\vspace{1mm}\noindent\textbf{Instance-based adversarial loss ---} The role of $D_{\text{inst}}$ is to provide adversarial cues on the realness of the separated sources without context. That is, $D_{\text{inst}}$ assesses the realness of each source individually:
	\begin{align*}
		[s_1] \ &/ \ [\hat{s}_1] \ \hdots \ [s_K] \ / \ [\hat{s}_K].
	\end{align*}
	Throughout the paper, we use brackets $[ ~ ]$ to denote the $D$'s input and left~/~right for \mbox{real~/~fake} separations (not division). 
	Hence, individual {real~/~fake} separations (instances) are input to $D_{\text{inst}}$, which learns to classify them as \mbox{real / fake (Fig.~\ref{fig:instance}).
	$D_\text{inst}$ is trained to maximize}
	\begin{align*}
		\mathcal{L}_{\text{inst}} = \frac{1}{K}  \sum_{k=1}^{K} 
		(\mathcal{L}_{\text{inst}}^{\text{real}, k} + 
		\mathcal{L}_{\text{inst}}^{\text{fake}, k}),
	\end{align*}
	where $\mathcal{L}_{\text{inst}}^{\text{real}, k}$ and $\mathcal{L}_{\text{inst}}^{\text{fake}, k}$ correspond to the hinge loss~\cite{lim2017geometric}:
	\begin{align*}
		\mathcal{L}_{\text{inst}}^{\text{real}, k} &= \min\left(0, -1 +  D_{\text{inst}}(s_{k})\right), \\ 
		\mathcal{L}_{\text{inst}}^{\text{fake}, k} &= \min\left(0, -1 - D_{\text{inst}} (\hat{s}_k)\right).
	\end{align*}
	Previous works also explored using $D_{\text{inst}}$. However, they used source specific setups where each $D_{\text{inst}}$ was specialized in a source type, e.g., for music source separation each $D_{\text{inst}}$ was specialized in bass, drums, and vocals~\cite{stoller2018adversarial,guso2022loss}, or for speech source separation $D_{\text{inst}}$ was specialized in speech \cite{subakan2018generative,chen2018permutation}. Yet, each $D_{\text{inst}}$ for universal sound separation is not specialized in any source type (are source agnostic) and assesses the realness of any audio, regardless of its source type. 
	
\begin{figure}[t]
	\centering
	\vspace{-1mm}
	\hspace{5mm}\includegraphics[width=0.88\linewidth]{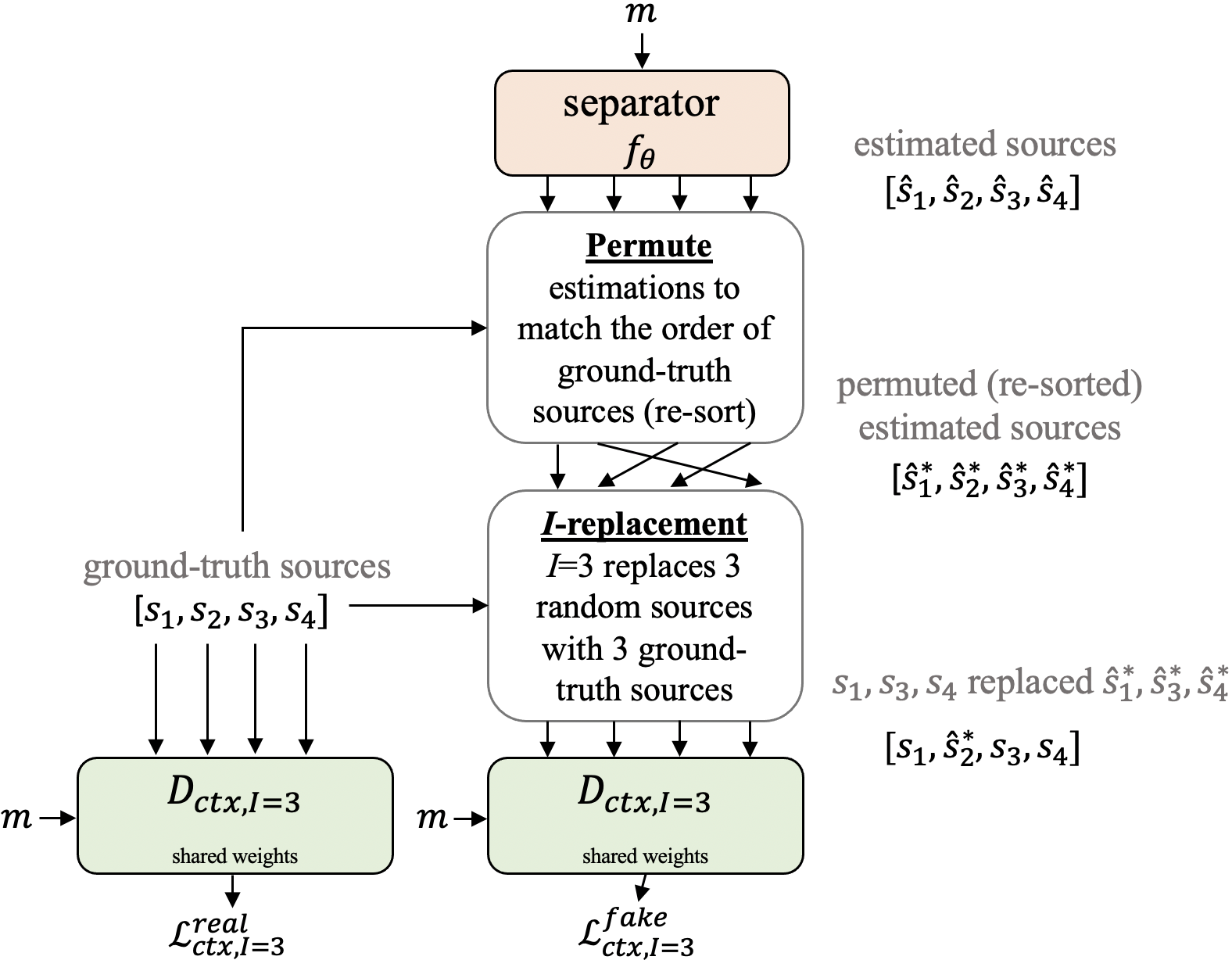}
	\vspace{-1mm}	
	\caption{\mbox{$I$-replacement context-based adversarial loss ($I=3$, $K = 4$).} }
	\label{fig:context}
	\vspace{-3mm}
\end{figure}

	\vspace{1mm}\noindent\textbf{$I$-replacement context-based adversarial loss ---} The role of $D_{\text{ctx}, I}$ is to provide adversarial cues on the realness of the separated sources considering all the sources in the mix (the context):
	\begin{align*}
		[s_1, \dots, s_K] \ &/ \ [\bar{s}_1, \dots, \bar{s}_K].
	\end{align*}
	Here, all the separations are input jointly to provide context to $D_{\text{ctx}, I}$, which learns to classify them as real / fake. $D_{\text{ctx}, I}$ can also be conditioned on the input mix $m$:
	\begin{align*}
		[m, s_1, \dots, s_K] \ &/ \ [m, \bar{s}_1, \dots, \bar{s}_K].
	\end{align*}
	Fake inputs contain entries $\bar{s}_{k}$, obtained by randomly getting $I<K$ source indices without replacement $\Lambda_I \subset \{1, \dots, K\}$, and {substituting the selected estimated sources $\hat{s}^*_k$ with their ground-truth $s_k$:}
	\begin{align}
		\label{eq:replacement}
		\bar{s}_k = \begin{cases}
			s_k & \text{if } k \in \Lambda_I, \\ 
			\hat{s}^*_k  & \text{otherwise},
		\end{cases}
	\end{align}
	where $\hat{s}^*_k = [\mathbf{P}^* \mathbf{\hat{s}}]_k $ and $\mathbf{P}^*$ is the optimal permutation matrix minimizing Eq.~\ref{eq:pit} with $\mathcal{L}$ as in Eq.~\ref{eq:loss}. 
	Note that finding the right permutation $\mathbf{P}^*$ is important to replace the selected source $\hat{s}^*_k$ with its corresponding ground-truth ${s}_k$, because the (source agnostic) estimations do not necessarily match the order of the ground-truth (see Fig.~\ref{fig:context}).
	Also, note that the $I=0$ case corresponds to the standard context-based adversarial loss used for speech source separation (without $I$-replacement, see Table~\ref{table:sota}). 
	Thus, our work generalizes adversarial PIT for universal sound separation by proposing a novel $I$-replacing schema that explicitly uses the ground-truth to guide the adversarial loss. 
	$D_{\text{ctx}, I}$ is trained to maximize
	\begin{align*}
		\mathcal{L}_{\text{ctx},I} = \mathcal{L}^{\text{real}}_{\text{ctx},I} + \mathcal{L}^{\text{fake}}_{\text{ctx},I}, 
	\end{align*}
	where we again use the hinge loss~\cite{lim2017geometric}:
	\begin{align*}
		\mathcal{L}^{\text{real}}_{\text{ctx},I} &= \min(0, - 1 + D_{\text{ctx}, I}(s_1, \dots, s_K)), \\ \mathcal{L}^{\text{fake}}_{\text{ctx},I} &= \min(0, - 1 - D_{\text{ctx}, I} (\bar{s}_1, \dots, \bar{s}_K)).
	\end{align*}

	\vspace{1mm}\noindent\textbf{Training with multiple discriminators ---} We have just presented $D_{\text{inst}}$ and $D_{\text{ctx}, I}$ in the waveform domain: $D^{\text{wave}}_{\text{inst}}$ and $D^{\text{wave}}_{\text{ctx},I}$. Next, we introduce them in the magnitude STFT ($D^{\text{STFT}}_{\text{inst}}, D^{\text{STFT}}_{\text{ctx},I}$) and mask ($D^{\text{mask}}_{\text{inst}}, D^{\text{mask}}_{\text{ctx},I}$) domains, and explain how to combine them. 
	The short-time Fourier transform (STFT) of a mix $m$ is defined as $M=\text{STFT}(m)$. The magnitude STFT is then obtained by taking the absolute value of $M$ in the complex domain, namely $\vert M \vert$. 
	Similarly, we denote as $\vert S_k \vert$ and $\vert \hat{S}_k\vert$ the magnitude STFTs of the target and the estimated sources, respectively. Ratio masks 
	\begin{align*}
				\label{eq:irm}
		R_k = \frac{ \vert S_k \vert}{\sum_{k=1}^{K} \vert S_k\vert} 
	\end{align*}
	are used to filter sources out from the mix by computing $S_k = M \odot R_k$, where $\odot$ denotes the element-wise product. Following the same notation as above, the input to the instance-based \mbox{$D_{\text{inst}}^{\text{STFT}}$ and $D_{\text{inst}}^{\text{mask}}$ is}
	\begin{equation*}	
		\begin{array}{cc}
			D^{\text{STFT}}_\text{inst}:  & 	 \left[\vert S_1\vert\right]\ / \ [\vert \hat{S}_1\vert ] \ \hdots \ \left[\vert S_K\vert \right]\ / \ [ \vert\hat{S}_K \vert], \\ 
			D^{\text{mask}}_\text{inst} : &  [R_1 ] \ / \ [ \hat{R}_1 ]  \ \hdots \  [R_K ] \ / \ [ \hat{R}_K ],
		\end{array}
		\vspace{-1mm}
	\end{equation*}
	and for the context-based $D_{\text{ctx,}I}^{\text{STFT}}$ and $D_{\text{ctx,}I}^{\text{mask}}$ (conditioned on $m$) is
		\vspace{-1mm}
	\begin{equation*}
		\begin{array}{cc}
			D^{\text{STFT}}_{\text{ctx}, I}:  & 	[ \vert M\vert  ,  \vert S_1\vert , \dots,  \vert S_K \vert ] \ / \ [ \vert M \vert,  \vert \bar{S}_1\vert , \dots,  \vert \bar{S}_K \vert ],  \\ 
			D^{\text{mask}}_{\text{ctx}, I} : &  [\vert M\vert  ,  R_1, \dots, R_K] \ / \ [\vert  M \vert, \bar{R}_1, \dots, \bar{R}_K ],
		\end{array}
	\end{equation*}
	where $\vert \bar{S}_k \vert$ and $\bar{R}_k$ entries follow the same $I$-replacement procedure as in Eq.~\ref{eq:replacement}. 
	Here, for $D_{\text{ctx},I}^{\text{STFT}}$ and $D_{\text{ctx},I}^{\text{mask}}$, the optimal permutation matrix  $\mathbf{P}^*$ required for re-sorting the fake examples is computed considering the L1-loss between magnitude STFTs or masks.
	The motivation behind combining multiple discriminators is to facilitate a richer set of adversarial loss cues to train $f_\theta$~\cite{guso2022loss,binkowski2019high,kumar2019melgan}, such that both $D_\text{inst}$ and $D_{\text{ctx},I}$ can provide different perspectives with respect to the same signal in various domains: waveform, magnitude STFT, and mask. 
	Hence, in addition to train each $D$ alone, one can train multiple combinations, e.g., $D^{\text{wave}}_{\text{inst}}+D^{\text{wave}}_{\text{ctx}, I}$, 
	$D^{\text{wave}}_{\text{ctx}, I}+D^{\text{STFT}}_{\text{ctx}, I}+D^{\text{mask}}_{\text{ctx}, I}$, or any combination of the discriminators above. 
	However, the more discriminators used, the more computationally expensive it is to run the loss, and the longer it takes to train $f_\mathbf{\theta}$ (but does not affect inference time).
	To the best of our knowledge, training with multiple discriminators \mbox{has never been considered for source separation before.}

	\vspace{1mm}\noindent\textbf{Separator loss ---} In adversarial training, $f_\theta$ is trained such that its (fake) separations $\mathbf{\hat{s}}$ are misclassified by the discriminator(s) as ground-truth ones  $\mathbf{{s}}$ (real). To do so, during every adversarial training step, we first update the discriminator(s) (without updating $f_\theta$) based on $\mathcal{L}_\text{inst}$, $\mathcal{L}_{\text{ctx},I}$, or any combination of the losses above. 
	Then, $\mathcal{L}_\text{sep}$ is minimized to train $f_\theta$ without updating the discriminator(s). For example, when using $D^{\text{wave}}_{\text{inst}}$ (with $D^{\text{wave}}_{\text{inst}}$ frozen) we minimize
	\vspace{-1.5mm}
	\begin{align*}
		\mathcal{L}_{\text{sep}} = -\frac{1}{K} \sum_{k=1}^{K} D^{\text{wave}}_{\text{inst}}(\hat{s}_k), 
	\end{align*}

		\vspace{-1.5mm}
		
	\noindent when using $D^{\text{STFT}}_{\text{ctx}, I}$ (with $D^{\text{STFT}}_{\text{ctx}, I}$ frozen) we minimize
	\begin{align*}
		\mathcal{L}_{\text{sep}} = - D^{\text{STFT}}_{\text{ctx}, I}(\vert \bar{S}_1\vert , \dots, \vert \bar{S}_K\vert ),
	\end{align*}

	\noindent or when using $D^{\text{STFT}}_{\text{inst}}$ and $D^{\text{STFT}}_{\text{ctx}, I}$ conditioned on $m$ (with $D^{\text{STFT}}_{\text{inst}}$ and $D^{\text{STFT}}_{\text{ctx}, I}$ frozen) we minimize
			\vspace{-2mm}
	\begin{align*}
		\mathcal{L}_{\text{sep}} = -\frac{1}{K} \sum_{k=1}^{K} & D^{\text{STFT}}_{\text{inst}}( \vert \hat{S}_k\vert )   - D^{\text{STFT}}_{\text{ctx}, I}(\vert  M \vert , \vert \bar{S}_1\vert , \dots, \vert \bar{S}_K\vert ).
	\end{align*}

	\vspace{-1.5mm}

	\noindent Again, note that we use the hinge loss~\cite{lim2017geometric}. 
	While we are not presenting all possible loss combinations for brevity, from the above examples one can infer all the {combinations we experiment with in section~\ref{discussion}.} Finally, we can also add an $\mathcal{L}_{\text{PIT}}$ term (as in Eqs.~\ref{eq:pit} and ~\ref{eq:loss}) to adversarial PIT: $\mathcal{L}_\text{sep} + \lambda \mathcal{L}_\text{PIT}$,
	where $\lambda$ scales $\mathcal{L}_\text{PIT}$ such that it is of the same magnitude as $\mathcal{L}_\text{sep}$ \cite{esser2021taming}. All previous adversarial PIT works for speech source separation used $\mathcal{L}_\text{sep} + \lambda \mathcal{L}_\text{PIT}$ {(Table~\ref{table:sota})}. Yet, in section~\ref{discussion} we show that our adversarial training setup allows dropping $\mathcal{L}_\text{PIT}$ while still obtaining competitive results, possibly because of the strong cues provided by $D_{\text{ctx}, I}$ (with $I$-replacement) and the multiple discriminators.
	To the best of our knowledge, we are the first to report results similar to $\mathcal{L}_\text{PIT}$ \mbox{with a purely adversarial setup (cf.~\cite{chen2018permutation}).}

	\section{EXPERIMENTAL SETUP}
	\label{setup}
	
	\noindent\textbf{Dataset, evaluation metrics, and baseline ---} We use the reverberant FUSS dataset, a common benchmark for universal sound separation with 20\,k / 1\,k / 1\,k (train / val / test) mixes of 10\,s with one to four sources~\cite{wisdom2021s, fonseca2017freesound, fonseca2021fsd50k}. 
	Metrics rely on the scale-invariant SNR~\cite{wisdom2021s}:
	\begin{align*}
		\hspace{-2mm}\text{SI-SNR}(s_k,\hat{{s}}_k^*) =  10 \log_{10} \frac{ \Vert {\alpha s_k} \Vert^2 + \epsilon}{\Vert \alpha s_k - \hat{{s}}_k^*\Vert^2 + \epsilon },  
	\end{align*}
	where $\alpha = \frac{{s_k}^T\hat{s}_k^*+\epsilon}{\Vert s_k \Vert ^2+ \epsilon}$, $\epsilon = 10^{-5}$, and $\hat{s}^*_k = [\mathbf{P}^* \mathbf{\hat{s}}]_k $ with  $\mathbf{P}^*$ being the optimal source-permutation matrix maximizing SI-SNR. 
	Further, to account for inactive sources, estimate-target pairs that have silent target 
	sources are discarded~\cite{wisdom2020unsupervised}. For mixes with one source, we compute $\text{SI-SNR}_\text{S} = \text{SI-SNR}(s_k, \hat{s}^*_k)$, which is equivalent to $\text{SI-SNR}(m, \hat{s}^*_k)$ since with one-source mixes the goal is to bypass the mix (the S sub-index stands for single-source\footnotemark[1]).
	For mixes with two to four sources, we report the average across sources of the $\text{SI-SNR}_\text{I}=\text{SI-SNR}(s_k, \hat{s}^*_k)-\text{SI-SNR}(s_k, m)$ 
	(the I sub-index stands for improvement\footnote{$\text{SI-SNR}_\text{S}$ is named as 1S or SS in~\cite{wisdom2021s,wisdom2020unsupervised} and $\text{SI-SNR}_\text{I}$ as MSi in~\cite{wisdom2021s,wisdom2020unsupervised}.}). 
	Note that we are using the standard SI-SNR formulation as in~\cite{wisdom2020unsupervised} instead of the alternative (less common) SI-SNR in~\cite{wisdom2021s,wisdom2021sparse}, and that we use the reverberant FUSS dataset as in~\cite{wisdom2021s,dcase} instead of its dry counterpart~\cite{wisdom2021sparse,wisdom2020unsupervised} because it is more realistic. As such, the results in~\cite{wisdom2021s,wisdom2020unsupervised,wisdom2021sparse} are not comparable with ours because they either use a different SI-SNR formulation or a different dataset.
	To compare our work against a meaningful state-of-the-art baseline we use the DCASE model, a TDCN++ predicting STFT masks~\cite{dcase}. It is trained on the reverberant FUSS dataset, and we evaluate it with the {metrics based on the standard SI-SNR.} Finally, we report $\text{SI-SNR}_\text{S}$ for consistency~\cite{wisdom2021s,wisdom2020unsupervised}, but $\text{SI-SNR}_\text{I}$ scores are more relevant for comparing models since most $\text{SI-SNR}_\text{S}$ scores are already very close to the upper-bound of 39.9\,dB (see Table~\ref{table:results}).

	\vspace{1mm}\noindent\textbf{Separator ---} 
	The mix $m$ of length $L=160000$ (10\,s at 16\,kHz) is mapped to the STFT domain $M$, with windows of 32\,ms and 25\% overlap (256~frequency bins and 1250~frames).  From $M$ we obtain its magnitude STFT $\vert M \vert$, that is input to a U-Net $g_\theta$ that predicts a ratio mask $\hat{R}_k$. The mask is obtained using a softmax layer $\sigma$ across the source dimension $k$: $\mathbf{\hat{R}}=\sigma(g_\theta(\vert M\vert))$,
	such that $\sum_{k} \vert{\hat{S}_k}\vert=\vert M\vert$. Then, we filter the estimated STFTs out of the mix with ${\hat{S}_k}={M \odot \hat{R}_k}$, and use the inverse STFT to get the waveform estimates: $\mathbf{\hat{s}} = f_\theta(m) = \text{iSTFT}( \mathbf{\hat{S}})$. 
	Hence, our separator can be trained in the waveform domain (with $\mathcal{L}_\text{PIT}$, $D_{\text{ctx}, I}^\text{wave}$, $D^{\text{wave}}_{\text{inst}}$), in the magnitude STFT domain (with $D_{\text{ctx}, I}^\text{STFT}$, $D^{\text{STFT}}_{\text{inst}}$), and/or in the mask domain (with $D_{\text{ctx}, I}^\text{mask}$, $D^{\text{mask}}_{\text{inst}}$).
	Our U-Net $g_\theta$ (of 46.9\,M parameters) consists of an encoder, a bottleneck, and a decoder whose inputs include the corresponding encoder's block outputs~\cite{ronneberger2015u,jansson2017singing}. The {encoder} is built of 4~ResNet blocks, each followed by a downsampler that is a 1D-CNN (kernel size=3, stride=2). The number of channels across encoder blocks is [256, 512, 512, 512]. The {bottleneck} consists of a ResNet block, self-attention, and another ResNet block (with all layers having 512 channels). The {decoder} block is built of 4~ResNet blocks, reversing the structure of the encoder, with upsamplers in place of downsamplers, reversing the number of channels of the encoder. The upsamplers perform linear interpolation followed by a 1D-CNN (kernel size=3, stride=1).  A final linear layer adapts the output to predict the expected number of sources ($K=4$).
	
	\vspace{1mm}\noindent\textbf{Discriminators ---} Each $D$ is of around 900\,k parameters, are fully convolutional, and output one scalar. $D^{\text{wave}}_{\text{inst}}$ and $D^{\text{wave}}_{\text{ctx},I}$ rely on a similar model: 4~1D-CNN layers (kernel size=4, stride=3), interleaved by LeakyReLUs, with the following number of channels: [$C$, 128, 256, 256, 512], where $C=1$ for $D^{\text{wave}}_{\text{inst}}$, and $C=5$ or $4$ for $D^{\text{wave}}_{\text{ctx}, I}$, depending if it is $m$-conditioned or not. Then, the 512 channels are projected to 1 with a 1D-CNN (kernel size=4, stride=1), and the final linear layer maps \mbox{the remaining temporal dimension into a scalar.}
	$D^{\text{STFT}}_{\text{inst}}$, $D^{\text{STFT}}_{\text{ctx},I}$, $D^{\text{mask}}_{\text{inst}}$, and $D^{\text{mask}}_{\text{ctx},I}$ are similar to $D^{\text{wave}}_{\text{inst}}$ and $D^{\text{wave}}_{\text{ctx},I}$, with the difference that 1D-CNNs are 2D (kernel size=4$\times$4, stride=3$\times$3) and the number of channels is [$C$, 64, 128, 128, 256].

	\vspace{1mm}\noindent\textbf{Training and evaluation setup ---} Models are trained until convergence (around 500\,k iterations) using the Adam optimizer, and the best model on the validation set is selected for evaluation. For training, we adjust the learning rate $\{ 10^{-5},10^{-4},10^{-3}\}$ and batch size $\{16, 32, 64, 96, 128\}$ such that all experiments, including ablations and baselines, get the best possible results. 
	Finally, we use a mixture-consistency projection~\cite{Wisdom_ICASSP2019} at inference time (not during training) because it systematically improved our $\text{SI-SNR}_\text{S}$ without degrading $\text{SI-SNR}_\text{I}$. 
	Our best model was trained for a month with 4~V100 GPUs with a learning rate of $10^{-4}$ and a batch size of 128. 
	
	
	\section{ EXPERIMENTS AND DISCUSSION}
	\label{discussion}
	
	First, in Table~\ref{table:ctxt}, we study various $D_{\text{ctx},I}$ configurations. We observe that the standard adversarial PIT ($I=0$, as in speech source separation) consistently obtains the worst results for universal sound separation. In contrast, the models trained with $I$-replacement ($I>0$) consistently obtain the best results. 
	We hypothesize that with $I=0$ $f_\theta$ does not separate much. Instead, it tends to approximate the naive solution of bypassing the mix. We can see this with the $\text{SI-SNR}_\text{S}$ scores, which tend to be closer (if not the same) to the $\text{SI-SNR}_\text{S}$ of the lower and upper bounds in Table~\ref{table:results}. 
	Overall, we note that the $I$-replacement context-based adversarial loss seems key to generalize adversarial PIT for universal sound separation, where multiple heterogeneous sources are separated. This contrasts with adversarial PIT works for speech source separation, where two similar sources are separated.
	We argue that the universal sound separation case is more challenging, as speech separation discriminators can judge the realness of separations based on speech cues, but discriminators for universal sound separation cannot as sources can be of any kind. 
	We hypothesize that the effectiveness of ${D}_{\text{ctx,}I>0}$ can be attributed to: 
	\begin{itemize}
		\item Replacing ${\hat{s}^*}_k$ with ${{s}}_k$ explicitly guides the adversarial loss to perform source separation. Note that $D_{\text{ctx}, I=0}$ (and $D_{\text{inst}}$) focuses on assessing the realness of its input. Under this setup, a naive solution is to always bypass the mix, which looks as real as one-source mixes where the goal is to bypass the mix.
		To avoid this naive solution, some guidance like the $I$-replacement is required. 
		\item It is more difficult for ${D}_{\text{ctx,}I>0}$ to distinguish between real and fake separations, because fake ones contain replacements. Consequently, such replacements help defining a non-trivial task for the discriminator that results in a better adversarial loss to train $f_\theta$.
	\end{itemize}

	\begin{table}[t]
		
		\begin{center}
			\resizebox{0.91\columnwidth}{!}{%
				\begin{tabular}{c | c@{\hskip 1.2mm}c | c@{\hskip 1.2mm}c | c@{\hskip 1.2mm}c }
					\toprule
					$m$ & $D_{\text{ctx}, I}^\text{STFT}$ & SI-SNR & $D_{\text{ctx}, I}^\text{mask}$ & SI-SNR & $D_{\text{ctx}, I}^\text{wave}$ &  SI-SNR \\				\midrule
					Yes & $I$$=$$0$  & ~4.9 / 35.7 & $I$$=$$0$   & ~5.9 / 35.8 & $I$$=$$0$  & ~8.5 / 39.9  \\ 
					Yes & $I$$=$$1$  & 11.0 / 33.2 & $I$$=$$1$   & ~9.3 / 33.6 & $I$$=$$1$  & 11.3 / 35.5  \\ 
					Yes & $I$$=$$2$  & 10.7 / 33.1 & $I$$=$$2$   & ~9.6 / 34.3 & $I$$=$$2$  & 11.9 / 36.3  \\ 
					Yes & $I$$=$$3$  & 11.7 / 33.4 & $I$$=$$3$   & ~8.6 / 32.6 & $I$$=$$3$  & 12.1 / 35.9  \\ 	\midrule
					No & $I$$=$$0$  & ~5.7 / 35.0 & $I$$=$$0$   & ~3.8 / 36.8 & $I$$=$$0$  & ~8.5 / 39.9  \\ 
					No & $I$$=$$1$  & ~9.9 / 32.8 & $I$$=$$1$   & ~6.7 / 34.1 & $I$$=$$1$  & 10.3 / 37.6 \\ 
					No & $I$$=$$2$  & 10.4 / 32.0  & $I$$=$$2$   & ~9.0 / 33.6 & $I$$=$$2$  & 11.0 / 36.0  \\ 
					No & $I$$=$$3$  & 10.2 / 31.0 & $I$$=$$3$   & ~9.7 / 33.5 & $I$$=$$3$  & 11.4 / 35.3  \\  \bottomrule	
				\end{tabular}
			}
		\end{center}
		\vspace{-4mm}
		\caption{Study of various $D_{\text{ctx}, I}$ configurations. $m$ column: $D_{\text{ctx}, I}$ is $m$-conditioned or not. SI-SNR column: \mbox{$\text{SI-SNR}_\text{I}$ / $\text{SI-SNR}_\text{S}$ in dB. }}
		\label{table:ctxt}
		\vspace{0mm}
		\begin{center}
			\resizebox{0.93\columnwidth}{!}{%
				\begin{tabular}{c@{\hskip 2mm}c@{\hskip 2mm}c@{\hskip 2mm}|c@{\hskip 2mm}c@{\hskip 2mm}c@{\hskip 2mm}|c@{\hskip 2mm}c@{\hskip 2mm}}
					\toprule
					$\mathcal{L}_{\text{ctx}, I}^\text{STFT}$ & $\mathcal{L}_{\text{ctx}, I}^\text{wave}$ & $\mathcal{L}_{\text{ctx}, I}^\text{mask}$   & $\mathcal{L}^{\text{STFT}}_{\text{inst}}$ & $\mathcal{L}^{\text{wave}}_{\text{inst}}$& $\mathcal{L}^{\text{mask}}_{\text{inst}}$ & $\mathcal{L}_\text{PIT}$ &  SI-SNR ($\uparrow$)	\\ \midrule
					\cmark&\cmark&\cmark&\cmark&\cmark&\cmark&\cmark& {13.5 / 37.2} \\ 
					\cmark&\cmark&\cmark&\cmark&\cmark&\cmark&-& 12.9 / 36.7  \\ 
					\cmark&\cmark&\cmark&-&-&-&-& {12.5 / 37.3} \\ 
					\cmark&\cmark&-&\cmark&\cmark&-&\cmark& \textbf{13.8} / 35.3\\ 
					\cmark&\cmark&-&\cmark&\cmark&-&-& 12.0 / 38.3\\
					\cmark&\cmark&-&-&-&-&-& {11.6 / 35.3}\\
					\cmark&-&-&-&-&-&-& {11.7 / 33.4} \\
					-&\cmark&-&-&-&-&-& 12.1 / 35.9  \\ 
					-&-&\cmark&-&-&-&-& ~8.6 / 32.6 \\
					-&-&-&\cmark&-&-&-& ~8.2 / 27.0 \\ 
					-&-&-&-&\cmark&-&-& ~{4.7 / 27.7} \\ 
					-&-&-&-&-&\cmark&-& ~4.5 / 36.8 \\
					-&-&-&-&-&-&\cmark&  12.4 / 37.4 \\ 
					\midrule
					\multicolumn{7}{l}{DCASE PIT baseline~\cite{dcase}}  & {12.8} / 37.5\\ 					
					\multicolumn{7}{l}{Lower-bound: return the input mix $m$}  & 0.0 / 39.9\\ 
					\multicolumn{7}{l}{Upper-bound: ideal ratio STFT masks~\cite{luo2019conv}}  & 25.3 / 39.9 \\  												 										
					\bottomrule
					
				\end{tabular}
			}
		\end{center}
		\vspace{-4mm}
		\caption{Comparison of adversarial PIT variants and baselines. SI-SNR column: $\text{SI-SNR}_\text{I}$ / $\text{SI-SNR}_\text{S}$ in dB. All $D_{\text{ctx}, I}$ above are $m$-conditioned with $I$$=$$3$, since it outperforms other setups (Table  \ref{table:ctxt}). All adversarial PIT ablations (rows 1-11 {\footnotesize \&} Table \ref{table:ctxt}) use the same $f_\theta$.}
		\vspace{-3mm}
		\label{table:results}

	\end{table}

	\noindent Next, we study the discriminators introduced in section~\ref{adversarialPIT} and their combination. In Table~\ref{table:ctxt}, we note that the $m$-conditioned $D_{\text{ctx}, I=3}$ generally outperform the rest. Hence, and for simplicity, in Table~\ref{table:results} we only experiment with this setup. 
	We note the following trends: 
	\begin{itemize}
		\item Our best result using adversarial PIT improves the state-of-the-art by 1\,dB (from 12.8 to 13.8\,dB) and improves the $\mathcal{L}_\text{PIT}$ baseline by 1.4\,dB (from 12.4 to 13.8\,dB). Informal listening also reveals that our best model separations more closely match the ground-truth sources and contain less {spectral holes} than the DCASE baseline (audio examples are available online\footnote[2]{\url{http://jordipons.me/apps/adversarialPIT/}}).
		{Spectral holes} are ubiquitous across mask-based source separation models, and are the unpleasant result of over-suppressing a source in a band where other sources are present. Adversarial training seems appropriate to tackle this issue since it improves the realness of the separations by avoiding spectral holes (which are not present in the training dataset).
		We also compare our {best model score (13.8\,dB) against the lower and upper bounds (25.3 and 0\,dB)} to see that there is still room for improvement {(also note this in our examples online\footnotemark[2]).}
		
		\item Our best results are obtained when combining multiple discriminators with $\mathcal{L}_\text{PIT}$ (over 13\,dB). This shows that complementing the adversarial loss with $\mathcal{L}_\text{PIT}$ is beneficial, and confirms that using multiple discriminators in various domains can help to improve the separations' quality. We also note that $\mathcal{L}_\text{PIT}$ alone and the best adversarially trained models (without $\mathcal{L}_\text{PIT}$) obtain similar results \mbox{(around 12.5\,dB).} 
		Hence, purely adversarial models can obtain comparable results to $\mathcal{L}_\text{PIT}$ alone even without explicitly optimizing for $\mathcal{L}_\text{PIT}$, in Eq.~\ref{eq:loss}, which is closely related to SI-SNR.

		\item When studying models trained with one $D$, we note that {$D_{\text{ctx}, I}$ alone tends to obtain better results than $D_{\text{inst}}$ alone.} We argue that the replacements in $D_{\text{ctx}, I}$ explicitly guide the separator to perform source separation, while for $D_{\text{inst}}$ this is not the case. In addition, $D_{\text{ctx}, I}^\text{wave}$ alone obtains a competitive score of 12.1\,dB, which can be improved up to 12.9\,dB if combined with 5~additional discriminators. Hence, results can be improved by using multiple discriminators but one can save computational resources by choosing the right $D$ without dramatically compromising the results. Finally, even though $D_{\text{inst}}$ alone under-performs the rest, we note that it can help improving the results when combined with $D_{\text{ctx}, I}$.
		
	\end{itemize}

	\section{CONCLUSIONS}
	\label{conclusion}

	We adapted adversarial PIT for universal sound separation with a novel $I$-replacement context-based adversarial loss and by training with multiple discriminators. With that, we improve the separations by 1.4\,dB $\text{SI-SNR}_\text{I}$ and reduce the unpleasant presence of spectral holes just by changing the loss, without changing the model or dataset. 
	Even with the improved results, we also stress that the obtained separations can still be improved by an important margin.

	\vspace{1mm}
	
	\section{ACKNOWLEDGMENTS}

	Work done during Emilian's internship~at~Dolby. Emilian's thesis is supported by the ERC grant no.~802554 (SPECGEO). Thanks to S. Wisdom and J. Hershey for your help with the metrics and baselines.

	\bibliographystyle{IEEEbib}
	\bibliography{mybib}

\end{document}